\documentclass[11pt]{article}
\usepackage{epsfig,amsfonts}
\usepackage[fleqn]{amsmath}
\usepackage{amsthm,amssymb}
\usepackage{graphicx}
\usepackage{hhline}
\usepackage{cite}
\usepackage{wasysym}
\usepackage{mathrsfs}

\def\be{\begin{equation}}
\def\ee{\end{equation}}
\def\bdm{\begin{displaymath}}
\def\edm{\end{displaymath}}
\def\bea{\begin{eqnarray}}
\def\eea{\end{eqnarray}}

\newcommand{\Ellipse}{\@chooseSymbol{'142}}

\newcommand{\rd}{\mbox{d}}
\newcommand{\ri}{\mbox{i}}
\newcommand{\re}{\mbox{e}}

\begin{document}

\begin{titlepage}

\begin{flushright}
LPTENS-05/13\\
\end{flushright}

\vspace{.8cm}
                                                                                               
\begin{center}
\begin{LARGE}
{\bf Dual Form of the Paperclip Model}

\end{LARGE}

\vspace{1.5cm}
 
\begin{large}
 
{\bf Sergei L. Lukyanov}$^{1,2}$ and 
{\bf Alexander B. Zamolodchikov}$^{1,2,3}$
 
\vspace{.2cm}

\end{large}
 
\vspace{.8cm}
 
{${}^{1}$ NHETC, Department of Physics and Astronomy\\
     Rutgers University\\
     Piscataway, NJ 08855-0849, USA\\
 
\vspace{.2cm}
 
${}^{2}$ L.D. Landau Institute for Theoretical Physics\\
  Chernogolovka, 142432, Russia

\vspace{.2cm}
and
\vspace{.2cm}
                                                                                                                          
${}^{3}$ Chaire Internationale de Recherche Blaise Pascal\\
Laboratoire de Physique Th${\acute {\rm e}}$orique
de l'Ecole Normale Sup${\acute {\rm e}}$rieure\\
24 rue Lhomond, Paris Cedex 05, France\\
}
\vspace{.4cm}

\end{center}                                                                                                                          
\begin{center}
\centerline{\bf Abstract} \vspace{.8cm}
\parbox{11cm}{
The ``paperclip model'' is 2D model of Quantum Field Theory with 
boundary interaction defined through a special constraint imposed
on the boundary values of massless bosonic fields (hep-th/0312168). 
Here we argue that this model admits equivalent ``dual'' 
description, where the boundary constraint is replaced by special 
interaction of the  boundary values of the bosonic fields with 
an additional  boundary degree of freedom. The dual form
involves the topological $\theta$-angle in explicit way.
}
                                                                                               
\end{center}                                                                                               
\vspace{.2cm}
\begin{flushleft}
\rule{4.1 in}{.007 in}\\
{October 2005}
\end{flushleft}
\vfil
                                                                                               
\end{titlepage}
\newpage

\section{Introduction}

In this work we describe the dual form of the ``paperclip model'' of 
boundary interaction in 2D Quantum Field Theory. The 
paperclip model was introduced in  Ref.\cite{LVZ}, where its
basic properties are discussed. The model involves two-component Bose field
${\bf X}(\sigma,\tau)=\big(X(\sigma,\tau)
, Y(\sigma,\tau)\big)$ which lives on a semi-infinite
cylinder with Cartesian coordinates $(\sigma,\tau)$,\   
$\sigma\geq 0,\ \tau\equiv\tau+1/T$\,\footnote{Euclidean formulation
of the theory is implied. 
Due to the compactification $\tau\equiv
\tau+1/T$ 
it is equivalent to the Matsubara representation of the $1+1$ dimensional
theory at thermodynamic equilibrium at the temperature $T$.}. 
In the bulk, i.e. at $\sigma>0$, the
dynamics is described by the free-field action:
\bea\label{baction}
{\mathscr   A}_{\rm bulk}[ X,Y ]=
{1\over 4\pi}\ \int_0^{1/T}\rd\tau\int_{0}^{\infty}\rd\sigma\
\big[\,
 {(\partial_{\nu} X)}^2+
(\partial_{\nu}  Y)^2\, \big]\ .
\eea
The interaction takes place at the boundary at $\sigma=0$, due to a 
non-linear boundary constraint: the boundary values 
${\bf X}_B = (X_B, Y_B) \equiv {\bf X}|_{\sigma=0}$ of the field ${\bf
  X}$ are restricted to the ``paperclip curve''
\bea\label{bconstraint}
r\,\cosh\big( {\textstyle{X_B\over {2\,b}}}\big) -
\cos\big(  {\textstyle{Y_B\over {2\, a}}}\big)= 0 \,, \ \ \qquad
|Y_{B}| \leq \pi a\ ,
\eea 
Here $a,\ b,$ and $r$ are real positive parameters, the first two
being related as follows\footnote{
The parameter $n$ used in \cite{LVZ}
is  
related to 
$a$ and $b$ as 
$$a={\textstyle{\sqrt {n+2}\over 2}}\, ,\ \ \  \
\ \ b={\textstyle {\sqrt {n}\over 2}}\ .
$$},
\bea\label{salskj}
a^2-b^2={\textstyle{1\over 2}}\ .
\eea

As usual, the
non-linear constraint requires renormalization, but one can check 
(up to two loops) that the RG transformation affects the curve
\eqref{bconstraint} only through renormalization of the parameter
$r$, which ``flows'' according to the equation
\bea\label{rgflow}
{E_*\over E} = 4b^2\  (1 - r^2)\, r^{4b^2}\,,
\eea
where $E$
is the
RG energy, and $E_{*}$ is the integration constant of
the RG equation, which sets up the ``physical scale'' in the model.
As in \cite{LVZ}, in what follows we always take the scale $E$ 
proportional 
to the temperature  $ T$, namely
\bea\label{kappadef}
E =  2\pi T\,.
\eea

It is useful to introduce also the external field
${\bf h} = (h_x,h_y)$ which couples to the boundary values ${\bf
  X}_B$, i.e. to add the boundary term
\bea\label{askj}
{\mathscr   A}_{\rm h}[X_B,Y_B]=\int_0^{1/T}\rd\tau \, (\, h_x\, X_B+
h_y\, Y_B\, )
\eea
to the action\ \eqref{baction}. Then, the first object of interest is
the partition function,
\bea\label{fint}
Z_0
=
\int\,{ \cal D} X { \cal D} Y\ 
\re^{-{\mathscr  A}_{\rm bulk}[X,Y]-
{\mathscr   A}_{\rm h}[X_B,Y_B] }
\,,
\eea
where the functional integration is over all fields ${\bf X}(\sigma,
\tau)$ obeying the boundary constraint\ \eqref{bconstraint}.

General definition of the model involves additional
parameter, the topological angle $\theta$.
Topologically, the paperclip curve \eqref{bconstraint} is a circle,
hence the configuration space for the field ${\bf X}(\sigma,\tau)$ consists of
sectors characterized by an integer $w$, the
number of times the boundary value ${\bf X}_B$ winds around the
paperclip curve when one  goes around the boundary at $\sigma=0$.
The contributions from the topological sectors can be weighted with
the factors $\re^{{\rm i}w\theta}$. Thus, in general
\bea\label{topsum}
Z_{\theta}= \sum_{w=-\infty}^{\infty}\re^{{\rm i}
w\theta}
 \ Z^{(w)}\, ,
\eea
where $Z^{(w)}$ is the functional integral \eqref{fint} taken over the
fields from the sector $w$ only. Physics of the model, in particular
its infrared (i.e. low temperature) behavior, depends on $\theta$ in
a significant way (see \cite{LVZ,LTZ} for details).
 
The ultraviolet (high-$T$) limit of the paperclip model is
understood in terms of the conformally invariant ``hairpin model'' 
of boundary interaction \cite{LVZ}. In this limit the parameter 
$r$ tends to zero, and the paperclip curve becomes a composition 
of two ``hairpins'', as shown in Fig.\,1. 

\begin{figure}[h]
\centering
\includegraphics[width=10cm]{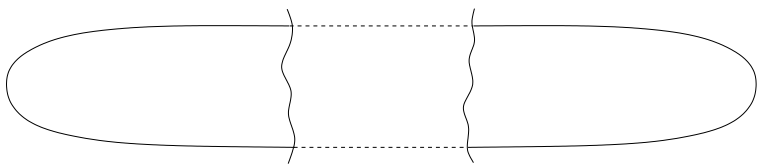}
\caption{The paperclip formed by junction of two hairpins.}
\label{fig-ampla}
\end{figure}

\noindent
The hairpin model is defined by replacing the paperclip constraint 
\eqref{bconstraint} by the non-compact ``hairpin'' 
curve
${\textstyle {r\over 2}}\,\exp\big(\pm {\textstyle{X_B\over 2\,b}}\big)=
\cos\big( {\textstyle{Y_B\over 2\,a}}\big)$.
We refer to this
model as the ``left'' or the ``right'' hairpin, depending on
the sign in the exponential; the two models are related by simple 
field transformation $(X,Y)\to (-X,Y)$). 
Note that the left hairpin corresponds to
the right part in Fig.\,2, and vice versa (just like
the way a human
brain is wired to the rest of the body).
The left (right) hairpin
model is conformally invariant with the linear dilaton $D({\bf X})=
{X\over 2b}$ 
$\big( D({\bf X})=-{X\over 2b}\big)$. More details on the hairpin model can be found in 
\cite{LVZ} and in  Section\,\ref{sectwo}
below.

The paperclip model has many features in common with the so-called
``sausage'' sigma model studied in Ref.\cite{falz}. The UV splitting
of the paperclip into the hairpins is analogous to the UV splitting
of the sausage into two semi-infinite ``cigars'' (see \cite{falz}). 
Like the sausage sigma model, the paperclip model seems to be integrable
at two values of the topological angle, $\theta=0$ and $\theta=\pi$. 
The sausage model is known to admit a dual description, where the 
non-trivial metric is replaced by certain potential (``tachion'') term
in the action \cite{fatea}. This analogy is one of the reasons to
expect that  a similar dual representation exists for the paperclip
model. The aim of this work is to introduce the dual representations
of both the hairpin model and the paperclip model.

In this paper we argue that the paperclip model is equivalent (or
``dual'') to another model of boundary interaction. The dual model
also involves a two-component Bose field $(X(\sigma,\tau),{\tilde
  Y}(\sigma,\tau))$ (where ${\tilde Y}$ is interpreted as the 
T-dual\footnote{The T-dual of the free massless field is defined as usual,
through the relations:
  $\partial_{\tau}{\tilde Y} = \ri\, \partial_{\sigma} Y$ and $
\partial_{\sigma}{\tilde Y} = -\ri\, \partial_{\tau} Y$.
\label{dual}}
 of $Y$) on the semi-infinite cylinder, which  
has  free-field dynamics in the bulk, and obeys 
{\it no constraint} at the boundary $\sigma=0$;
instead it interacts with an additional boundary degree of
freedom. It is best to discuss the dual model in terms of its 
Hamiltonian representation, with the cyclic coordinate $\tau\equiv
\tau+1/T$ taken as the Euclidean (or Matsubara) time. In this picture 
the partition 
function \eqref{topsum} admits the dual representation as the 
trace
\bea\label{tracce}
Z_{\theta} = {\rm Tr}_{{\tilde{\cal H}}}\, \Big[\, 
\re^{-{{\hat H}\over T}}\, \Big]
\,,
\eea
taken over the space ${\tilde{\cal H}} = {\cal
  H}_{X,{\tilde Y}}\otimes {\mathbb C}^2$, where ${\cal H}_{X,{\tilde
    Y}}$ is the space of states of the two-component boson
$\big(X(\sigma), {\tilde Y}(\sigma)\big)$ on the half-line $\sigma
\geq 0$ (with no constraint at $\sigma=0$) and ${\mathbb C}^2$
is the two-dimensional space representing the new boundary
degree of freedom. The dual Hamiltonian in \eqref{tracce} consists of 
the bulk and the boundary parts, ${\hat H} = 
{\hat H}_{\rm bulk} + {\hat H}_{\rm boundary}$.  The bulk part is just the 
free-field Hamiltonian
\bea\label{hbulk}
{\hat H}_{\rm bulk} ={1\over 4\pi}\,
 \int_{0}^{\infty}\rd\sigma \,\big[\, \Pi_{X}^2 + \Pi_{\tilde
  Y}^2 + (\partial_{\sigma} X)^2 + (\partial_\sigma {\tilde
  Y})^2\, \big]\ ,
\eea
where $(\Pi_{X}, \Pi_{\tilde Y})$ are momenta
conjugated to the field operators  $(X, {\tilde
  Y})$ 
\footnote{Normalization 
is such that, for instance, $[\,  X (\sigma)\, ,\, \Pi_X(\sigma')
\,]=
2\pi\ri\, \delta(\sigma-\sigma')\ .$} acting 
in ${\cal H}_{X,{\tilde Y}}$ (${\hat H}_{\rm bulk}$
acts as identity in the ${\mathbb C}^2$ component of ${\tilde{\cal H}}$). The
boundary term describes coupling of the boundary values $(X_B,
{\tilde Y}_B)\equiv 
(X,{\tilde Y})|_{\sigma=0}$ of the fields  to the additional boundary
degree of freedom represented by ${\mathbb C}^2$ ($\sigma_{\pm}$ and
$\sigma_3$ are the Pauli matrices acting in ${\mathbb C}^2$),
\bea\label{hbdry}
{\hat H}_{\rm boundary} = h_x\,X_B + \pi a\,h_y\,\sigma_3 +  {\hat
  V}\ ,
\eea
where\footnote{We assume canonical conformal normalization
of the boundary vertex operators 
(see e.g. Eq.\eqref{ope}) involved in this interaction term.
}:
\bea\label{shasyaasd}
{\hat V}=\mu_{B}\ \Big[\,
 \sigma_+ \cosh\big(\, b X_B-\ri\, {\textstyle{\theta\over 2}}\, \big)\
\re^{{\rm i}a  {\tilde Y}_B}+
 \sigma_-
\cosh\big(\,b X_B+\ri\, {\textstyle{\theta\over 2}}\, \big)\
\re^{ -{\rm i}a  {\tilde Y}_B}\, \Big]\ ,
\eea
with $\mu_B$ related to the scale $E_*$ in \eqref{rgflow} as
\bea\label{alskjs}
\mu_{B}=\sqrt{2E_*\over \pi}\ .
\eea
In Eqs.\eqref{hbdry},\,\eqref{shasyaasd}, 
 $a,\, b$ and $(h_x, h_y)$ are the same as in
\eqref{bconstraint},\,\eqref{askj}.

\section{Dual form of the hairpin \label{sectwo}}

In this section we describe the dual form of the conformal hairpin
model. To be definite, throughout this section we concentrate attention
on the {\it left} hairpin model; the right hairpin is obtained by
reflection $X \to -X$. The left hairpin boundary constraint has
the form
\bea\label{hairpin}
 {\textstyle{r_{*}\over 2}}\  
\exp\big({\textstyle{X_B\over {2\,b}}}\big)=
\cos\big(  {\textstyle{Y_B\over {2\,a} }}\big)\ ,  \ \ \qquad
|Y_{B}| \leq \pi a\ . \eea 
Here $r_{*}$ relates to the energy scale \eqref{kappadef} in a simple
way 
\bea\label{rstar}
{E_*\over 2\pi T}=4b^2\, (r_{*})^{4 b^2}\ .
\eea

The boundary state of the (left) hairpin model was described in \cite{LVZ},
\bea\label{hairstate}
\langle\, B_{\supset}\, | = \int_{\bf P}\rd^2{\bf P}\ 
B_{\supset}({\bf P})\ \langle\,I_{\bf
  P}\, |\ ,
\eea
where $\langle\, I_{\bf P}\, |$ are the Ishibashi states associated
with the $W$-algebra of the hairpin model (see \cite{LVZ} and 
Section\,\ref{sectwoone} below), ${\bf P}$ is
the zero-mode momentum
of the free boson ${\bf X}$, and the amplitude $B_{\supset}({\bf
  P})$ has the following explicit form in terms of the components 
$(P,Q)$ of the vector ${\bf P}$
\bea\label{bright}
B_{\supset}(P,Q) = {\rm g}_D^2\ r_*^{-2{\rm i} b  P}\ 
\ {2b\ \Gamma(2\ri b\,P)\, \Gamma\big(1+
\ri\,  {P\over 2b}\big)\over\Gamma\big({1\over 2} -
 a Q+ \ri b P\big)\,
\Gamma\big({1\over 2} + a Q
+  \ri bP
\big)}\ ,
\eea
where ${\rm g}_D=2^{-1/4}$.
The amplitude $B_{\supset}(P,Q)$ coincides with the
partition function of the hairpin model on the semi-infinite cylinder
of circumference $1/T$:
\bea\label{zetb}
Z_{\supset}(h_x,h_y) =
B_{\supset}(P,Q)\,,
\eea
where the dependence of $Z_{\supset}$ on the parameters $h_x,\,h_y$
is brought about through the linear boundary term \eqref{askj}, and 
$P,Q$ in the right hand side are related to the these parameters as
follows
\bea\label{pxi}
P=\ri\,  { h_x\over T}\, ,\ \ \ \ Q=\ri\,  { h_y\over T}\ .
\eea

\subsection{$W$-algebra and dual potential \label{sectwoone}}

As was explained in \cite{LVZ}, the hairpin boundary condition
is conformally invariant (with linear dilaton $X/b$), and moreover it
has extended conformal symmetry with respect to certain $W$-algebra.
Here we use the notation $W_{\supset}$ for the $W$-algebra of the 
left hairpin model. It is generated by a set of 
local currents $\{\,W_s (z), s=2,\,3,\,4\ldots\,\}$ (built from
derivatives of the  free 
fields $X,Y$) which commute with two ``screening operators''\footnote{Note
that we write the screening operators in terms of the T-dual field
${\tilde Y}$.
This is done in preparation to the discussion of the dual
hairpin below. At this point the distinction makes no difference 
in the definition of the
$W_{\supset}$-algebra, since the equation \eqref{screenings} is
sensitive to the holomorphic parts of the fields $X$ and $Y$ only.},
\bea\label{screenings}
\oint_z\rd w \ W_s (z) \ \re^{bX \pm {\rm i}a{\tilde Y}}(w,{\bar w})  =0\ .
\eea
Then, it is almost trivial observation that this $W$-symmetry is present
in any model of boundary interaction which has no boundary constraint, 
but instead whose action has an additional boundary potential term
\bea\label{bpotential}
{\mathscr  A}_{\rm BP} = \int_{0}^{1/T}\rd\tau\ \Big[\, S_{+}\,\re^{bX_B
    +{\rm i}a{\tilde Y}_B} + S_{-}\,\re^{bX_B -{\rm i}a{\tilde
      Y}_B}\, \Big]\ .
\eea
Here $S_{\pm}$ may be either c-numbers, or more generally any
non-trivial boundary degrees of freedom whose own dynamics is 
``topologically invariant'', i.e. invariant with respect to any 
diffeomorphism $\tau \to f(\tau)$ of the boundary. Since such
boundary interaction has the hairpin $W$-algebra as its symmetry, it
seems natural to expect that under appropriate choice of the boundary degree
of freedom ${S}_{\pm}$ the model with the boundary potential 
\eqref{bpotential} is equivalent (or, in modern speak, ``dual'') to
the left hairpin model. 

Because of the essentially quantum nature of the boundary degree of
freedom $S_{\pm}$ (see below), it will be convenient to discuss in
terms of the Hamiltonian representation of the model, as was mentioned 
in Introduction. The partition function of the left hairpin model
is written as the trace  ${\rm Tr}\big[\,\re^{-{{\hat H}\over T}}\,\big]$, where 
${\hat H}$ is the sum ${\hat H}_{\rm bulk} + {\hat V}_{\supset}$ of
the bulk free-field Hamiltonian \eqref{hbulk}, and the boundary term
\bea\label{hairdualv}
{\hat V}_{\supset} = {\mathbb S}_{+}\ 
\re^{b X_B+{\rm i}a {\tilde Y}_B}+
{\mathbb S}_{-}\ \re^{b X_B-{\rm i} a {\tilde Y}_B}\, 
,\eea
corresponding to the term \eqref{bpotential} in the action. Here 
${\mathbb S}_{\pm}$ are the operators associated with the 
boundary degree of freedom $S_{\pm}(\tau)$ in \eqref{bpotential}\,. 
The latter operators must commute with the field operators, i.e. 
$[\, {\mathbb S}_{\pm}\, ,\,  {\bf X}(\sigma)]=
[\, {\mathbb S}_{\pm}\,,\, {\bf\Pi}(\sigma)\,] =0$ at
any $\sigma$,  
lest the $W$-symmetry of the theory be violated, but the commutation 
relations among the boundary observables ${\mathbb S}_{\pm}$ themselves 
are not fixed {\it a priori}. Our goal here is to identify the  algebra 
of these operators in the dual hairpin model, as well as its 
representation $\rho$.

Some relations can be inferred from the following simple argument. 
The boundary potential term \eqref{bpotential} vanishes in the limit 
$X \to -\infty$. In the absence of this term both fields $X$ and 
${\tilde Y}$ would obey the von Neumann (free) boundary conditions. 
Equivalently, in the absence of the potential term the field $Y$ would 
obey the Dirichlet (fixed) boundary condition, i.e. the boundary
values $(X_B,Y_B)$ would lay on a straight brane parallel to the 
$X$-axis. Note that in the same limit $X\to-\infty$ the right 
hairpin curve \eqref{hairpin} becomes a composition of two parallel
branes,
\bea\label{braneass}
Y_B\to \pm \pi a\ \ \ {\rm as}\ \  X_B\to -\infty\, ,
\eea
separated by the distance $2\pi a$. In the full hairpin curve, these
two ``legs'' are bridged at the right, allowing for a passage from 
the upper leg to the lower leg and vice versa. In the dual
representation of the hairpin model such passages should be 
attributed to two terms in the
boundary potential \eqref{bpotential}. The boundary vertex operators 
$\re^{\pm {\rm i}a {\tilde Y}_B}(\tau)$ create jumps in the boundary 
value of $Y$ exactly of the desired magnitude, $Y_B (\tau+0) - 
Y_{B}(\tau-0) = \pm 2\pi a$. The fact that the hairpin has only
two ``legs'' \eqref{braneass} clearly suggests that the operators 
${\mathbb S}_{\pm}$ in \eqref{hairdualv} must obey the ``fermionic''
relations\footnote{It is important at this point that our definition of the hairpin (and of the
paperclip) model involves {\it uncompactified} field $Y$. Also, the bound
$|Y_B| < \pi a$ in \eqref{hairpin} (and in \eqref{bconstraint}) is essential.
Without the bound,  Eq.\eqref{hairpin} (as well as Eq.\eqref{bconstraint})
would define a series of disconnected curves, $Y \to Y+ 4\pi a\, {\mathbb Z}$
copies of the original hairpin (or paperclip). Although the models of
boundary interaction which involve more then one copy also deserve attention, they
are different from the hairpin (paperclip) model as defined in \cite{LVZ},
and we do not address them here.}
\bea\label{fermionic}
{\mathbb S}_{+}^2={\mathbb S}_{-}^2=0\ .
\eea 

So far in the discussion of the dual hairpin we have ignored the coupling
\eqref{askj} to the external field $(h_x,h_y)$. Adding the term
$h_x X_B$ to the dual hairpin Hamiltonian is
straightforward, but in the description in terms of the T-dual field 
${\tilde Y}$ the term $h_y Y_B$ would be non-local. One can circumvent
this difficulty by introducing the ``fermion number'' operator ${\mathbb
  N}$, associated with the boundary degrees of freedom ${\mathbb
  S}_{\pm}$, which satisfies with them the following commutation
relations:
\bea\label{fnumber}
[\, {\mathbb N}\, ,\, {\mathbb S}_{+}\, ]=2\, {\mathbb S}_{+}\, ,
\ \ \ \ \ \
[\, {\mathbb N}\, ,\, {\mathbb S}_{-}\, ]=-2\, {\mathbb S}_{-}\  .
\eea
Then one can check that the sum $Y_B - \pi a\,{\mathbb N}$  commutes
with the Hamiltonian \eqref{hairdualv}. Therefore, in the presence of
the external field ${\bf h}$, expected form of the full Hamiltonian
of the dual hairpin model is
\bea\label{fullhairdual}
{\hat H}= {\hat H}_0 + {\hat V}_{\supset}\,,
\eea
where ${\hat V}_{\supset}$ is the boundary potential operator
\eqref{hairdualv}, and
\bea\label{hnot}
{\hat H}_{0} = {\hat H}_{\rm bulk} + h_x\,X_B + \pi
a\,h_y\ {\mathbb N}\,,
\eea
with the operators ${\mathbb N}$ and ${\mathbb S}_{\pm}$ in
\eqref{hairdualv} satisfying the relations \eqref{fermionic} and \eqref{fnumber}.

\subsection{Singularities of the hairpin  boundary amplitude}

Looking at the right hairpin amplitude \eqref{bright} as a function
of complex variables $P$ and $Q$, one observes two sets of
singularities. First one 
is a sequence of poles at $2\ri b\,P =0,\, -1,\, -2,\ldots$ due
to the first gamma-factor in the numerator in \eqref{bright}. These
poles admit straightforward interpretation in terms of potential
divergences of the functional integral \eqref{fint}, with the
non-compact boundary constraint \eqref{hairpin}, at the infinite end
of the hairpin $X \to -\infty$ (see Ref.\cite{LVZ}). The second set is a string of poles
at
\bea\label{poles}
P_k = 2\ri b\,k\,, \ \ \ \ \ \ k= 1,\, 2,\,3\ldots
\eea
due to the second gamma-factor in \eqref{bright}. 
Note that in the weak-coupling limit of the
hairpin model, i.e. at $b\to\infty$, these poles depart to infinity. 
Clearly, in terms of the hairpin functional integral these singularities
represent non-perturbative effects. Instead, the poles \eqref{poles}
become most visible at small $b$, which is the weak coupling domain in
the dual representation of the hairpin model, and indeed they admit 
simple interpretation in terms of the dual hairpin model. Since the 
boundary potential \eqref{bpotential} vanishes in the limit $X\to
-\infty$, there is a potential divergence of the functional integral 
associated with the dual 
hairpin\footnote{
It is not difficult to write down the full functional integral for the dual
hairpin model, which involves integration over the fields $X$ and
${\tilde Y}$, as well as the integral over the boundary spin ${\bf S} =
(S_{+},S_{-}, S_{3})$ with the Wess-Zumino term (see, e.g., Ref.\cite{Polyakov}). Such
expression is not very useful for our analysis, except for the observation
that the only terms in the full action which involve the constant mode
of $X$ are the boundary potential term \eqref{bpotential}, and the term
${\tilde {\mathscr A}}_{{\bf h}} = \int_{0}^{1/T}\rd\tau\,\big[\, 
h_x X_B(\tau) + \pi a h_y\,S_3
(\tau)\, \big]$ responsible for the
coupling to the external field $(h_x,h_y)$.
}.
Following \cite{Goulian}, one can 
first integrate out the constant mode of the field $X$. This integration
produces poles in $P$ exactly at the points \eqref{poles}
whose residues 
\bea\label{resus}
R_k =\ri\ 
{\rm Res}_{ P= 2{\rm i}
 bk}\big[\,  Z_{\supset}\big(-\ri PT, -\ri QT\big)\, \big]\ \ \ 
\ \ (k=0,\,1,\,2\ldots)
\eea
are expressed through the integrals of the $2k$-point correlation functions
\bea\label{intcorr}
R_k =(2\pi T)^{nk^2}\ 
 {\cal T}\int \rd\tau_{2k}\cdots \rd\tau_{1}\ \langle\langle \,{\hat V}_{\supset}
(\tau_{2k})\cdots {\hat V}_{\supset}(\tau_1)\, \rangle\rangle_{0}\ ,
\eea
where the $\tau$-ordering symbol ${\cal T}$ signifies that the integration is
performed over the domain $1/T\geq\tau_{2k}\geq\cdots \geq \tau_{1}
\geq 0$. In \eqref{intcorr} ${\hat V}_{\supset} (\tau) = 
\re^{\tau{\hat H}_0} \,{\hat V}_{\supset}\, \re^{-\tau{\hat
    H}_0}$ is the unperturbed Matsubara operator associated with
the boundary potential \eqref{hairdualv}, and $\langle\langle\, \cdots
\, \rangle\rangle_{0} 
\equiv {\rm Tr}\big[\,\cdots\,\re^{-{{{\hat H}_0}\over T}}\, \big]$. 
Since the 
unperturbed Hamiltonian ${\hat H}_0$ involves no interaction between the boundary
variables ${\mathbb S}_{\pm}$ and the fields $X,\,{\tilde Y}$, the expectation
value in \eqref{intcorr} factorizes in terms of these two parts of the 
system. In view of \eqref{hairdualv}, it can be written in the form
\bea\label{formfactor}
R_k =&&\sum_{\epsilon_{i}=\pm}
{\rm  Tr}_{\rho}
\big[\, {\mathbb S}_{\epsilon_{2k}}\,{\mathbb S}_{\epsilon_{2k-1}}\,
\cdots{\mathbb S}_{\epsilon_{1}}\,\re^{{\rm i}\pi
  aQ\,{\mathbb N}}\,\big]\times \\
&& \ \ \ \ \ \ \ {\cal T}\int \rd\tau_{2k} \cdots \rd\tau_{1}\ \langle\, 
{\cal V}^{\epsilon_{2k}}_{B}(\tau_{2k})\,\cdots \,{\cal
  V}^{\epsilon_{1}}_{B}(\tau_{1})
\,\rangle_{0}\ ,\nonumber
\eea
which involves the traces over the space of states $\rho$ of the 
boundary degrees of freedom ${\mathbb S}_{\pm}$, as well as the
free-field expectation values $\langle\,\cdots\,\rangle_{0}$ 
of the boundary values ${\cal V}^{\pm}_B (\tau)$ of the vertex 
operators \eqref{screenings}. Thanks to the relations
\eqref{fermionic}, there are only two nonvanishing contributions to
the sum in \eqref{formfactor}, with $(\epsilon_1\,
\cdots\,\epsilon_{2k}) = (+ - + - \cdots + -)$ and 
$(\epsilon_1\,\cdots\,\epsilon_{2k}) = (- + - +  \cdots- +)$. With this
observation, and using explicit form of the free-field correlators 
in \eqref{formfactor}, this expression can be brought to the form:
\bea\label{factorform}
R_{0} &=& 
{{\rm g}_{D}^2\over{2\pi}}\ {\rm Tr}_{\rho}\big[\, 
\re^{{\rm i}\pi a Q {\mathbb N}}\, \big]\ ,
\nonumber\\
R_{k} &=& {{\rm g}_{D}^2\over{2\pi}}\ (2\pi T)^{-k}\,\,F_k\,\,G_k\ \ \ \ \ \ \ (k=1,\,
2,\,\ldots)\ ,
\eea
where
\bea\label{xitraces}
F_k = \re^{{\rm i}\pi a Q}\ 
{\rm Tr}_{\rho}\big[\, \re^{{\rm i}\pi a Q {\mathbb N}}\,
({\mathbb S}_{-}{\mathbb S}_{+})^k\,  \big]=
\re^{-{\rm i}\pi a Q}\
{\rm Tr}_{\rho}\big[\, \re^{{\rm i}\pi a Q {\mathbb N}}\,
({\mathbb S}_+ {\mathbb S}_-)^k\,  \big]\ ,
\eea
and  $G_k$ are given by the 
$2k$-fold integrals
\bea\label{gk}
&&G_k=\int_{0}^{2\pi }{\rd u_k}\int_{0}^{u_k }{\rd v_k}
\int_{0}^{v_k }{\rd u_{k-1}}\cdots
\int_{0}^{v_{2} }{\rd u_1}\int_{0}^{u_1 }{\rd v_1}\nonumber\times\\
&&\ \ \ \ \ \  
\prod_{j>i}^k \Big[ 4 \sin\big({\textstyle{u_j-u_i\over 2}}\big)
\sin\big({\textstyle{v_j-v_i\over 2}}\big)\Big]\
\prod_{j\geq i}^k\Big[2 \sin\big({\textstyle{u_j-v_i\over 2}}\big)
\Big]^{-4b^2-1}\times\nonumber\\ && \ \ \ \ \ \  
\prod_{j> i}^k\Big[2 \sin\big({\textstyle{v_j-u_i\over 2}}\big)
\Big]^{-4b^2-1}
\ \  2\ \cos\Big[\, 
a Q\big( \pi+\sum_{i=1}^k(v_i-u_i)\big)\, \Big]\ .
\eea
The overall factor
${{\rm g}_{D}^2\over 2\pi}$ in \eqref{factorform} appears because
\eqref{formfactor} involves unnormalized free-field correlation
functions; here ${\rm g}_D = 2^{-1/4}$ is well known ``boundary entropy''
factor\ \cite{affleck} associated with the Dirichlet and von-Neumann boundary
conditions for a free boson\footnote{The $g$-factor of uncompactified
boson $X$ with the von-Neumann boundary condition diverges. Formally,
it involves the factor ${{\rm g_D}\over 2\pi}\ {\rm d} X_0$, where $X_0$ is the 
zero mode of $X$\ \cite{saleur}. In \eqref{intcorr} the integration over $ X_0$ is 
already performed -- this integration was the origin of the poles
\eqref{poles}.
Additional factor 
${\rm g}_D$ comes from the unperturbed partition function of
${\tilde Y}$. Since it  is the T-duality transform of $Y$, its partition
function equals to
$\int \rd{\tilde Q} \ \langle \,B_N\,|\, {\tilde Q}\,\rangle = {\rm g}_D$,
where $\langle\, B_N\, |$ is the boundary state associated with the
von Neumann boundary condition for the field ${\tilde Y}$.
}.

Obviously, the way they are written above, the integrals \eqref{gk}
diverge for all positive $b^2$. As is common in conformal perturbation
theory, we assume here a version of ``analytic regularization'', where
the expressions are understood as analytic continuations of these
integrals from the domain $\Re e\,b^2 <0$\,\footnote{More generally,
these divergences have to be canceled by adding a counterterm 
$M\,\re^{2bX_B}$  (with the cutoff-dependent
coefficient $M$) to the Hamiltonian \eqref{hnot}.}.
This procedure is performed  in  Appendix. Remarkably, the integrals 
\eqref{gk} are evaluated in a closed form (the calculations are
presented in  Appendix),
\bea\label{gkanswer}
G_k= {(2\pi)^{k+1} \ (-4 b^2)^{-k}\ \Gamma(1-4 b^2 k)\over
k!\, \Gamma\big(\, {1\over 2}-2b^2\, k-a Q\, \big)\
\Gamma\big({1\over 2}- 2b^2\, k+a Q\, \big)}\ \  \ \ \ 
(k\geq 1)\ .
\eea
Note that the $Q$-dependence of \eqref{gkanswer} is exactly as
expected from \eqref{bright}, and moreover \eqref{factorform} 
coincide with the residues of \eqref{zetb},\,\eqref{bright} at the 
points \eqref{poles} provided
\bea\label{trzero}
{\rm Tr}_{\rho}\big[\, \re^{{\rm i}\pi a  Q {\mathbb N}}\, \big]
=2\, \cos(\pi a Q)\ ,
\eea
and
\bea\label{trk}
{\rm Tr}_{\rho}\big[\,
({\mathbb S}_{-}{\mathbb S}_{+})^k\,  \big]=
{\rm Tr}_{\rho}\big[\,
({{\mathbb S}_{+}}{\mathbb S}_{-})^k\,  \big]=
\big({\textstyle{E_*\over 2\pi}}\big)^k
\ \ \ \ \ (\, k=1,\,2\ldots\, )\ .
\eea

The equations \eqref{trzero} and \eqref{trk} are sufficient to
identify the representation $\rho$ of the algebra 
\eqref{fermionic},\,\eqref{fnumber} 
of the boundary degrees of freedom. From
\eqref{trzero} one finds
\bea\label{dimrho}
{\rm dim}(\rho) =2\ ,
\eea
and
\bea\label{trn}
{\mathbb N}^2 = 
{\mathbb I}\ , \ \ \ \ \ {\rm Tr}_{\rho}\big[\, {\mathbb N}\, \big] = 0\ .
\eea
Next, the two-dimensional representation of the algebra must satisfy
additional relation
\bea\label{anticomm}
\{\, {\mathbb S}_{+}\, ,\, {\mathbb S}_{-}\,\}={\textstyle{E_*\over 2\pi}}
     \,\times \, {\mathbb I}\ .
\eea
Indeed, $\rho$ is necessarily an irreducible representation of the 
algebra. As follows from Eqs.\eqref{fermionic},\,\eqref{fnumber},
the anticommutator
$\{\, {\mathbb S}_{+}\, ,\, {\mathbb S}_{-}\}$ commutes
with ${\mathbb S}_{\pm}$ and ${\mathbb N}$, therefore
it should be a constant in $\rho$.
With Eq.\eqref{trk} this implies the condition\ \eqref{anticomm}.

There is a unique (up to equivalence) two-dimensional representation of \ 
\eqref{fermionic},\,\eqref{fnumber},\,\eqref{anticomm} which satisfies
\eqref{trn}:
\bea\label{sigmarep}
\rho\ :\ \  {\mathbb S}_{+}=\sqrt{\textstyle{E_*\over 2\pi}}
\ \, \sigma_+\, ,\ \ \ \
{\mathbb S}_{-}=\sqrt{\textstyle{E_*\over 2\pi}}\ \, \sigma_-\, ,\ \ \
{\mathbb  N}=\sigma_3\ ,
\eea
where $\sigma_a$ are conventional Pauli matrices. Thus we identify the
boundary degree of freedom of the dual hairpin model with the spin
$s=1/2$. Note that two eigenvalues of $\sigma_3$ are associated with 
two legs of the hairpin. Note also that according to \eqref{sigmarep}
the operators ${\mathbb S}_{\pm}$ have dimension $[\,{energy}\, ]^{1\over
  2}$, as required by the balance of dimensions in
Eq.\eqref{bpotential} (in view of \eqref{salskj} the vertex operators
$\re^{bX_B\pm {\rm i}a{\tilde Y}_B}$ have dimensions $[\, {energy}\,]^{1\over
  2}$\,).
 
\section{Dual to the paperclip model}

The dual representation for the
paperclip model can be identified using similar  line 
of  arguments. The idea of the paperclip \eqref{bconstraint} being
the composition of the left and right hairpins makes it natural to
look for the Hamiltonian of the dual paperclip model in the
form
\bea\label{dualclip}
{\hat H} = {\hat H}_{\rm bulk}+h_x\, X_B+
 \pi a\,  h_y \,{\mathbb N} + {\hat V} \ ,
\eea
where ${\hat H}_{\rm bulk}$ is the same free-field bulk
Hamiltonian\ \eqref{hbulk}, and 
\bea\label{dualclipv}
{\hat V}  &= &\sqrt{\textstyle{E_*\over 2\pi}}\ \ 
\Big[\,{\mathbb A}_{+}\,
\re^{b X_B+
{\rm i}a {\tilde Y}_B}+
 {\mathbb A}_{-}\,
\re^{b  
X_B-{\rm i}a {\tilde Y}_B}+\ \ \ \ \ \ \ \nonumber\\
&&\ \ \ \ \ \ \ \ \ \ \ \ \ {\mathbb B}_{+}\, 
\re^{-b X_B+{\rm i} a {\tilde Y}_B}+
{\mathbb B}_{-}\,
\re^{-b X_B-{\rm i} a
{\tilde Y}_B}\, \Big]\ .
\eea
Here ${\mathbb A}_{\pm}$, ${\mathbb B}_{\pm}$ and
${\mathbb N}$ are operators representing boundary degrees of freedom,
which commute with the field operators ${\bf X}(\sigma), \ {\bf
  \Pi}(\sigma)$. Note that we have explicitly put the factor $\propto \sqrt{E_*}$ in
\eqref{dualclipv}, so that the
operators ${\mathbb A}_{\pm}$, ${\mathbb B}_{\pm}$ are dimensionless.
Obviously, the first two terms in \eqref{dualclipv}
are associated with the left  hairpin component of the
paperclip, i.e. the operators ${\mathbb A}_{\pm}$ play the same role
as 
${\mathbb S}_{\pm}/\sqrt{E_{*}\over 2\pi}$ in \eqref{hairdualv}. The last two 
terms in \eqref{dualclipv} are associated in a similar way with the
dual form of the right hairpin. This correspondence suggests that
the operators ${\mathbb B}_{\pm}$, as well as ${\mathbb A}_{\pm}$,
satisfy the fermionic relations analogous to \eqref{fermionic}
\bea\label{ffermionic}
{\mathbb A}_{\pm}^2 =0\ , \ \ \ \ \ \ \ \ {\mathbb B}_{\pm}^2 =0\ .
\eea
In addition, they have to satisfy the commutation relations with 
${\mathbb N}$
\bea\label{ffnumber}
[\, {\mathbb N}\, ,\, {\mathbb A}_{\pm}\,] = 
\pm 2\,{\mathbb A}_{\pm}\ , \ \ \ \ 
\ \ \ \ \ \ \ [\, {\mathbb N}\, ,\, {\mathbb B}_{\pm}\,] = 
\pm2\,{\mathbb B}_{\pm}\ ,
\eea
and the anticommutation relations
\bea\label{anticomma}
\{\, {\mathbb A}_{+}\, ,\,  {\mathbb A}_{-}\,\} =
 \{\, {\mathbb B}_{+}\, ,\, {\mathbb
  B}_{-}\, \} = {\mathbb I}\ .
\eea

Intuitively, these relations can be advocated by the 
same arguments that were considered in the previous section. The 
paperclip curve in  Fig.\,1 can be regarded as the combination of 
two nearly straight parallel D-branes connected to each other by the 
left and the right hairpin curves. Then, the presence of the vertex
operators $\re^{bX_B \pm {\rm i}a{\tilde Y}_B}$ in \eqref{dualclipv} is the 
way how the dual representation reflects the possibility of passages from
one straight brane to another via the  connection at the
right, i.e. at
sufficiently large positive  $X_B$. Likewise, the operators 
$\re^{-b
  X_B \pm {\rm i}a {\tilde Y}_B}$, which become significant at large
negative $X_B$, describe the transitions between the nearly parallel
branes via the connection at the left. This picture suggests that the space of
states associated with the boundary degrees of freedom in
\eqref{dualclip}, i.e. the supporting space of the representation
$\rho$ of the above algebra, is two-dimensional, with two basic 
vectors (the eigenvectors of ${\mathbb N}$) corresponding to the two 
constituent straight branes. Then,  Eq.\eqref{ffermionic} simply
express the statement that there is s single copy of the paperclip
curve (as is specified by the $Y_B$ bound in Eq.\eqref{bconstraint}). 
Also, since $\rho$ is irreducible, and since in the limit $E_{*}\to 0$ 
we have to recover \eqref{anticomm}, the relations \eqref{anticomma} 
follow. 

It is easy to check that any two-dimensional representation
$\rho$ of the algebra
\eqref{ffermionic},\,\eqref{ffnumber} and
\eqref{anticomma} is equivalent to the
following one:
\bea\label{rhoo}
\rho: \qquad {\mathbb A}_{\pm} = \re^{\pm{\ri}\, {\theta\over 2}}\ 
\sigma_{\pm}\ ,\ \ \ \ \ \ \ \ 
{\mathbb B}_{\pm } = \re^{\mp {\ri}\, {\theta\over 2}}\ \sigma_{\pm}\  ,
\eea
where $\theta$ is an arbitrary complex parameter.
It is possible to show that this  parameter must be real and, moreover,
it coincides with the $\theta$-angle of the paperclip model.

The easiest way to verify this identification of $\theta$  is to analyze specific
logarithmic divergences generated by the interaction
\eqref{dualclipv}. The divergences appear due to the singular term
in the Operator Product Expansions
\bea\label{ope}
\re^{bX_B\pm {\rm i}a{\tilde Y}_B}(\tau) \,\re^{-bX_B \mp {\rm i}a {\tilde
    Y}_B}(\tau') = {1\over{|\tau-\tau'|}} + {\rm regular\
  terms}\ ,
\eea
and it is easy to check that in view of \eqref{rhoo} they can be
absorbed by local boundary counterterm
\bea\label{counterterm}
-2E_*\cos(\theta)\ \log(\Lambda/E_{*})\ \int {\rd\tau\over 2\pi} \ ,
\eea
where $\Lambda$ is the  UV cut-off. Exactly the same counterterm, with
$\theta$ being the topological angle, is required in the original
formulation of the paperclip model, where its role
is to compensate for ``small instanton'' divergences (see \cite{LVZ}
for details).

The above observation suggests that the instanton contributions of the
original paperclip model are reproduced by certain terms of conformal
perturbation theory  of the dual 
model \eqref{dualclip},\,\eqref{dualclipv}. 
To be sure, the conformal
perturbation theory, understood as a regular
expansion in powers of $E_{*}$, can not be literally valid for this
model. The model has many properties in common with the boundary 
sinh-Gordon model, and in view of the analysis in \cite{bsinh}, one 
expects rather complicated structure of the UV expansion in
\eqref{dualclip}. For example, in the case $h_x=0$ and at
$T\gg E_{*}$ the partition function $Z(h_x,h_y\,|\, T)|_{h_x=0}$ of the model 
\eqref{dualclip} is expected to develop a large logarithmic term,
\bea\label{lskjsal}
Z(0,h_y\, |\, T) = G(
Q\, |\, \kappa)
\ \log\big(
 {\textstyle {1\over \kappa}}\big) + 
F( Q\, |\, \kappa)\ ,
\eea
which derives from large 
$\big(\sim b^{-1}\log({ 1\over \kappa}) \big)
$ fluctuations of the zero mode of the field
$X$. The coefficient $G$ depends on $Q=\ri\, { h_y
\over T}$ and 
\bea\label{kshsakhs}
\kappa =
{ {E_{*}\over 2\pi T}}\ .
\eea
Up to an overall factor $\kappa^{2\kappa\cos(\theta)}$ (whose origin
could be traced down to the singular term in Eq.\eqref{ope}),
it admits a small-$\kappa$ expansion in 
double series in powers of $\kappa$ and $\kappa^{1\over 2{b^2}}$ 
(the term $F$ has a similar expansion). 
The term $\sim\kappa$ of this expansion is given by the
integral
\bea\label{cpt}
G( Q\, |\, \kappa ) &=& {{{\rm g}_D^2}\over{2\pi b}}\,
{\rm Tr}_{\rho}\Big[\, 
\re^{{\rm i}\pi aQ{\mathbb N}}\ \Big\{ 1 +\\
&&{\cal T} \int\rd\tau_2\, \rd\tau_1\ \langle \,{
    V}_{\supset}(\tau_2)\,{\hat V}_{\subset}(\tau_1) + {\hat
    V}_{\subset}(\tau_2)\,{\hat V}_{\supset}(\tau_1)\,\rangle_0 
+ \ldots\, \Big\}\, \Big]\,, 
\nonumber
\eea   
where ${\hat V}_{\supset}$ and ${\hat V}_{\subset}$ correspond to the 
first two terms and the second two terms on Eq.\eqref{dualclipv},
respectively. Putting in explicit correlation functions of the
exponential fields in \eqref{dualclipv} and evaluating the trace,
one can bring this expression to a more explicit form
\bea\label{cptt}
&&G( Q \,|\, \kappa ) = {{\rm g}_D^2
\over{2\pi b}}\ \, 2\,\cos(\pi a
Q)\  \big(\,
1 + \kappa\, D(Q)\, \cos(\theta)+
\ldots\, \big)\, ,
\eea
where the factor  $\cos(\theta)$  in the second term appears as the 
result of evaluation of the trace using  Eqs.\eqref{rhoo},
and $D(Q)$ is the integral
\bea\label{alss}
D(Q) ={1\over 2\pi}
\  \int_{0}^{2\pi}{\rd u_2}\int_{0}^{u_2}{\rd u}_1\
{\cos\big( a Q (\pi-u_2+u_1)\big)
\over  \cos(\pi a
Q)\, \sin({u_2-u_1\over 2})}+{\rm counterterm}\, ,
\eea
where $u_1$ and $u_2$ differ from $\tau_1$ and $\tau_2$ in \eqref{cpt}
by a factor of $ 2\pi T$. The integral logarithmically diverges when
$u_1 \to u_2$, but its divergent part  cancels with 
the counterterm \eqref{counterterm}, and the
$Q$-dependent finite part is evaluated explicitly, in terms of
Euler's function  $\psi(x)={{\rm d}\over {\rm d} x}\, \log\Gamma(x)$. 
As the result, the expansion of $G(Q\, |\,\kappa)$ has the form
\eqref{cptt} with
\bea\label{cpttt}
D(Q) = 2\, \log(\kappa) - \psi
\big({\textstyle{1\over 2}}-a Q\big)-
\psi\big({\textstyle{1\over 2}}+a Q\big)\ .
\eea 
With this, the coefficient in front of $\cos(\theta)$ in \eqref{cptt}
exactly matches one-instanton contribution to the paperclip partition
function (see Eq.(110) of \cite{LVZ}).

\section*{Acknowledgments}
                                                                                                  
The authors are   grateful to  Vladimir V. Bazhanov,
Vladimir A. Fateev and Alexei B. Zamolodchikov 
for discussions and interest to this work. 

\bigskip
                                                                                                  
\noindent
The research is supported
in part by DOE grant $\#$DE-FG02-96 ER 40959.
ABZ gratefully acknowledges kind hospitality of Laboratoire de 
Physique Th\'eorique de l'Ecole Normale Sup\'erieure, and generous 
support from Foundation de l'Ecole Normale Sup${\acute {\rm e}}$rieure,
via  chaire Internationale de Recherche Blaise Pascal.

\bigskip
\bigskip

\section{Appendix: Calculation of the integral $G_k$\ \eqref{gk}}

\subsection{Contour integral representation of $G_k$}
 
Up to overall factor, the integrand in \eqref{gk} coincides with the
free-field expectation value of product of $2k$ chiral vertex operators
\bea\label{asisal}
V_{\pm}=
\re^{2b X_R\pm 2{\rm i}a  Y_R}\ ,
\eea
where ${\bf X}_R=(X_R,\, Y_R)$ in the exponent stands for the
holomorphic part of the Bose field ${\bf X}(\sigma,\tau)=
{\bf X}_R(\tau+\ri\sigma)+{\bf X}_L(\tau-\ri\sigma)$. As usual, the
free-field expectation values are fully determined by the two-point
functions,
\bea\label{twopoint}
\langle\,X_R (\tau) X_R(\tau')\,\rangle_{0} =
\langle\,Y_R (\tau) Y_R(\tau')\,\rangle_{0} = -{\textstyle{1\over 2}}\
\log\big[\sin\big(\pi T\, (\tau-\tau')\big)\,\big]\ .
\eea
                                                                                                   
It is useful to introduce the following set of integrated products
\bea\label{jahsh}
J_{p}(\epsilon_1 \cdots\epsilon_p )=
\int_{\tau_0}^{\tau_0+1/T }\rd \tau_p\int_{\tau_0}^{\tau_{p}}
\rd \tau_{p-1}\cdots
\int_{\tau_0}^{\tau_{2}}\rd \tau_{1}\
V_{\epsilon_p}(\tau_p)\cdots V_{\epsilon_1}(\tau_{1})\ ,
\eea
where $(\epsilon_1 \cdots\epsilon_p )$ is a set of signs, $\tau_0$ is
fixed real parameter, and integration is along the
real axis. Clearly, the integral \eqref{gk} is certain linear
combinations of expectation values of $J_{2k}( \epsilon_1
\cdots\epsilon_{2k})$ with $(\epsilon_1 \cdots\epsilon_{2k} )$ being
one of two alternating sign sequences $(+-+\cdots+-)$ and
$(-+-\cdots-+)$.
                                                                                                   
As the  first step in  calculation  we
transform \eqref{jahsh}\ into  contour  integrals.
To do this we introduce the ``screening operators'',
the integrals of the chiral vertex operators:
\bea\label{salksas}
x_{\pm}={1\over q-q^{-1}}\
\int_{\tau_0}^{\tau_0+1/T}
{\rd \tau}\ \re^{2b X_R\pm 2{\rm i}a  Y_R}\ ,
\eea
where\footnote{As it  is known, the
screening operators $x_{\pm}$, together with
the two zero mode operators, $\int_{\tau_0}^{\tau_0+1/T} {\rd \tau}\,
\big(\, a\partial_\tau X_R\pm \ri\, b\partial_\tau Y_R\, \big)$, 
form the Chevelley
basis of the Borel subalgebra of quantum superalgebra
$U_{q,q}(sl(2|1))$ \cite{Zhang}.}
\bea\label{ajhsgx}
q=\re^{ {\rm i}\pi n}\ .
\eea
Here and bellow in this Appendix  we use the notation from Ref.\cite{LVZ}:
\bea\label{ssjsl}
n\equiv 4\, b^2\ .
\eea
The  following relations,
\bea\label{klash}
&&(q-q^{-1})\, x_+J_{2k-1}(-\cdots -)=
J_{2k}(+\cdots -)-q^{k}\, J_{2k}(-\cdots +) \nonumber\\
&&(q-q^{-1})\, x_-J_{2k-1}(-\cdots -)=0 \nonumber\\
&&(q-q^{-1})\, x_+J_{2k}(-\cdots+)=J_{2k+1}(+\cdots +)\\
&&(q-q^{-1})\, x_-J_{2k}(-\cdots +)=q^k\ J_{2k+1}(-\cdots-)\ ,
 \nonumber
\eea
can be easily established by rearranging the integration domains.
Additional set of relations is obtained from these
by replacing $x_{\pm} \to x_{\mp}$ and
simultaneously changing all signs, $J_{p}(\epsilon_1\cdots \epsilon_p)\to
J_{p}(-\epsilon_1\cdots -\epsilon_p)$. Recursively applying 
\eqref{klash}, one can prove that
\bea\label{hsgahg}
&&J_{2k}(+\cdots -)=
(q-q^{-1})^k\ {(-1)^k\,  q^{-{k(k+1)\over 2}}\over  [k]_q!}\
\big(\, q^k\, (x_-x_+)^k+(x_+x_-)^k\, \big)\nonumber\\
&&J_{2k}(-\cdots +)=
(q-q^{-1})^k\ {(-1)^k\,  q^{-{k(k+1)\over 2}}\over  [k]_q!}\
\big(\, q^k\, (x_+x_-)^k+(x_-x_+)^k\, \big)\ ,\nonumber
\\
\eea
where the standard notations,
\bea\label{skasjh}
[k]_q!=[1]_q[2]_q\ldots[k]_q\, ,\ \ \ \ \ \
[k]_q={q^k-q^{-k}\over q-q^{-1}}\ ,
\eea
are used.
                                                                                                   
\begin{figure}[ht]
\centering
\includegraphics[width=4cm]{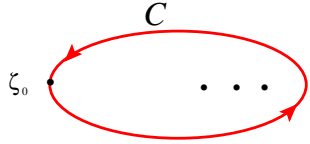}
\caption{The  integration
contour in Eq.\eqref{akkjs}.}
\label{fig-analita1}
\end{figure}

\begin{figure}[ht]
\centering
\includegraphics[width=6cm]{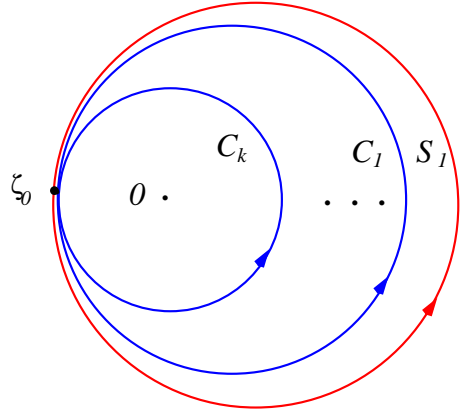}
\caption{The integration contours in Eq.\eqref{sssks}.}
\label{fig-count}
\end{figure}

To make the next step more transparent we change to the new coordinate
\bea\label{slkssk}
z=\re^{2\pi{\rm i}T \tau }\ .
\eea
The screening charges \eqref{salksas} become contour integrals in
the variable $z$. More precisely, the action of the operators
$x_{\pm}$ on any state created by a set of local insertions in the
$z$-plane is written as the integral
\bea\label{akkjs}
x_{\pm} (\,\ldots\,) =
{1\over q-q^{-1}}\ \int_{C} dz\ V_{\pm}(z)\, (\,\ldots\, )\ ,
\eea
where the contour $C$ starts from the point $\zeta_0 = \re^{2\pi{\rm
    i}T \tau_0 }$, goes around all the insertions in the
counterclockwise direction, and then returns back to $\zeta_0$, as is
shown in Fig.\,\ref{fig-analita1}. The
integrand in \eqref{akkjs} should be understood in terms of free-field
operator product expansions. In general, the contour is not closed
since operator product in the integrand is multivalued function of
$z$.
Using\ \eqref{hsgahg}\ the integral $G_k$ can be
rewritten in the form:
\bea\label{saaaks}
&&G_k=2\ (-1)^k \ q^{ k^2\over 2}\
{(q-q^{-1})^k\over
[k]_q!}\
\Big[\, \cos\big(\pi (a Q+
{\textstyle{nk\over 2}})\big)\times\\ &&
\ \ \  \langle\,  (x_+x_-)^k\, V_\lambda(0)\,  \rangle+
\cos\big(\pi (a Q-
{\textstyle{nk\over 2}})\big)\,
\langle\,  (x_-x_+)^k\,
V_\lambda(0)\,
 \rangle\, \Big]\, ,
\nonumber
\eea
or more explicitly
\bea\label{sssks}
&&G_k={2^{1-k}  \,\ri^k\, \re^{-{{\rm i}\pi\over 2}\,  nk^2} \over
\prod_{j=1}^k\sin(\pi n j)}\
\int_{C_k}\rd\zeta_k\int_{S_k}\rd z_k\ldots\int_{C_1}\rd\zeta_1
\int_{S_1}\rd z_1\times \nonumber \\ &&
\Big[\, \cos\big(\pi (a Q+
{\textstyle{nk\over 2}})\big)
\ \langle\, 
V_-(\zeta_k) V_+(z_{k})\ldots V_{-}(\zeta_1) V_+(z_1) V_\lambda(0)\, \rangle+\\
&&\ \  \ \ \cos\big(\pi (a Q-
{\textstyle{nk\over 2}})\big)\
\langle\, V_+(\zeta_k) V_-(z_{k})\ldots V_{+}(\zeta_1) V_-(z_1) V_\lambda(0)
\, \rangle\, \Big]\ ,
\nonumber
\eea
where the integration is over a set of contours $C_j$, $S_j$ 
$(j=1,2,\, \ldots\, k)$ each starting at $\zeta_0$ and returning to the same
point after going around the point $0$, and arranged so that
$C_k$ 
lays entirely (except for the point $\zeta_0$ itself) inside $S_k$,
$S_{k}$ lays inside $C_{k-1}$, $C_{k-1}$ lays inside $S_{k-1}$, etc.,
as is depicted in Fig.\,\ref{fig-count}.
The expectation values in the integrand in  \eqref{sssks} involves
a certain vertex operator,
\bea\label{vlambda}
V_{\lambda} =\exp\Big(-{\textstyle {n(\lambda_++\lambda_-)\over 2\, b}}
\, X_R+\ri\ {\textstyle{n(\lambda_--\lambda_+)\over 2\, a}}\, 
Y_R\, \Big)\ ,
\eea
with $\lambda_{\pm}$ to be specified below, inserted at the origin
$z=0$,  so that
\bea\label{kasjshs} &&
\langle\,
V_-(\zeta_k) V_+(z_{k})\ldots V_+(z_1) V_\lambda(0)\, \rangle=\ri^{-k}\,
\re^{-{{\rm i}\pi\over 2} n k^2}\,
\prod_{j=1}^{k} z_j^{n\lambda_+}
\zeta_j^{n\lambda_-}\times\\ &&
\prod_{j>i}\big[(z_j-z_i)(\zeta_i-\zeta_j)\big]\,
\prod_{j>i}\big[(\zeta_j-z_i)
 (z_j-\zeta_i)\big]^{-n-1} \,  \prod_{j=1}^k(\zeta_j-z_j)^{-n-1} .
\nonumber
\eea
We assume that the brunches of the power functions in \eqref{kasjshs}
are chosen to give real positive values at
\bea\label{slksjsasl}
0<z_1<\zeta_1\ldots<z_k<\zeta_k\ .
\eea
In Eqs.\eqref{vlambda},\,\eqref{kasjshs} the following notations are used
\bea\label{alksjalksj}
n\lambda_+={kn-1\over 2}+a Q+\varepsilon\, ,\ \ \ \ \
n\lambda_-={kn-1\over 2}-a Q\ ,
\eea
where  $\varepsilon$ is a complex parameter, which is assumed to be
small, and eventually will be sent to zero.
                                                                                                   
\subsection{Combinatorics of the contour integrals}
                                                                                                   
Now let us introduce another set of integrated products of vertex
 operators,
\bea\label{ajahsh}
I_{p}(\epsilon_1 \cdots\epsilon_p )=\int_{\zeta_0}
^{0 }\rd z_p
\int_{\zeta_0}^{z_p}\rd z_{p-1}\cdots
\int_{\zeta_0}^{z_{2}}\rd z_{1}
V_{\epsilon_1}(z_1)\cdots V_{\epsilon_p}(z_{p})V_{\lambda}(0)\ .
\eea
Here $V_{\lambda}$ is the vertex operator \eqref{vlambda}. Although
 the final result of  the calculations below does not depend on a choice
of the point $\zeta_0$, for convenience we choose it to be real and
 negative, and we assume that all the integrations in \eqref{ajahsh}
 are along the real axis. The operator products in \eqref{ajahsh} are
multivalued functions of the integration
 variables, and we assume the same choice of the branch as in
 \eqref{kasjshs}. 
Note that the
integrals \eqref{ajahsh} are similar but different from
\eqref{jahsh}, the main difference being in the form of 
integration contours.

\begin{figure}[ht]
\centering
\includegraphics[width=4cm]{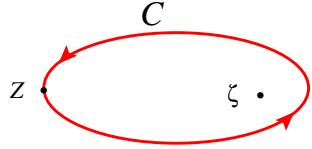}
\caption{The  contour of  analytic continuation
in Eq.\eqref{jhsagshg}.}
\label{fig-analit}
\end{figure}

The monodromies of the operator products in \eqref{ajahsh} are
determined by symbolic relations
\bea\label{jhsagshg}
{\cal A}_C[\, V_{\pm}(z)V_{\lambda}(\zeta)\, ]=q^{2\lambda_{\pm}} \
V_{\pm}(z)V_{\lambda}(\zeta)\ ,
\eea
and, in particular,
\bea\label{kajsk}
{\cal A}_C[\, V_{+}(z)V_{-}(\zeta)\, ]=q^{-2} \
V_{+}(z)V_{-}(\zeta)\ ,
\eea
where the symbol ${\cal A}_C[\ldots]$ denotes analytic continuation
in the variable $z$ along the contour $C$ shown on
Fig.\,\ref{fig-analit}.
Using these relations one can derive the following identities:
\bea\label{aslskjsl}
x_{+} I_{2k}(+\ldots -)&=&-q^{\lambda_+-k}\ [\lambda_+]_q\
I_{2k+1}(+\ldots +)
\nonumber \\
x_{+} I_{2k}(-\ldots +)&=&
-q^{\lambda_+-k}\, [\lambda_+-k]_q\, I_{2k+1}(+\ldots +)
\nonumber \\
x_{+} I_{2k-1}(+\ldots +)&=& 0
\\
x_{+} I_{2k-1}(-\ldots -)&=&
-q^{\lambda_+-k}\, [\lambda_+-k]_q\,
I_{2k}(+\ldots -)
-\nonumber \\ &&q^{\lambda_+-k}\, [\lambda_+]_q\, I_{2k}(-\ldots +)
\ ,\nonumber
\eea
as well as similar identities obtained from these by simultaneous
change of the signs: $\{x_{\pm}\to x_{\mp},\ \lambda_{\pm}\to
\lambda_{\mp},\ I_{p}(\epsilon_1\cdots \epsilon_p)\to
I_{p}(-\epsilon_1\cdots -\epsilon_p)\}$. Here the action of the
screening charges $x_{\pm}$ is defined by Eq.\eqref{akkjs}.
The recursion\ \eqref{aslskjsl}\ allows one
to express   the
ordered integrals\ \eqref{ajahsh}\  in terms of the screening
charges. In particular, one finds
\bea\label{skasjhsj}
I_{2k}(-\ldots+)={1\over c_k\, [\lambda_+]_q}\
\big(\, [\lambda_+-k]_q
\, (x_-x_+)^k+[\lambda_+]_q\,
 (x_+x_-)^k\, \big)\, V_{\lambda}(\zeta)\ ,
\eea
where
\bea\label{alksjska}
c_k=(-1)^k\
q^{k(\lambda_++\lambda_--k)}\ {[k]_q!\, [\lambda_++\lambda_--1]_q!\over
[\lambda_++\lambda_--k-1]_q!}\ .
\eea
                                                                                                   
Now assume that $\lambda_{\pm}$ are given by \eqref{alksjalksj}, and
consider the limit $\varepsilon\to  0$. It is easy to check that at
$\varepsilon=0$ the expression \eqref{ajahsh} becomes invariant with
respect to simultaneous rescaling of all the integration variables,
hence the integral develops logarithmic divergence at $z_i \to 0$. Thus,
as the function of $\varepsilon$, $I_{2k}(-+ \cdots +)$ is expected to
have a simple pole at $\varepsilon=0$. The appearance of the pole is very
explicit in \eqref{skasjhsj}, since the coefficient $c_k$ vanishes at
$\varepsilon=0$,
\bea\label{alsas}
c_k \to  \varepsilon  \  {2 \pi\,\ri
\over q-q^{-1}}\ (-1)^k\ [k]_q!\,[k-1]_{q}!\ .
\eea
The representation \eqref{skasjhsj} also  makes it easy to isolate the
residue at this pole,
\bea\label{alksjl}
&&\lim_{\varepsilon\to 0}\varepsilon\, I_{2k}(-\ldots+)=
{(-1)^k\, (q-q^{-1})\over 2\pi\ri\, [k]_q!\,[k-1]_q!\,
  \cos\big(\pi (a Q+{kn\over 2})\big)}
\times\\ &&\ \ \
\Big(\, \cos\big({\textstyle \pi (a Q-{kn\over 2}}\big)\
(x_-x_+)^k+\cos\big({\textstyle \pi (a Q+{kn\over 2}}\big)\
(x_+x_-)^k\, \Big)\, V_{\lambda}(\zeta)\, .\nonumber
\eea
Comparing\ \eqref{alksjl}\ with \eqref{saaaks} we observe that the
desired integrals $G_k$ can be expressed through such residues,
\bea\label{laskjss}
G_k&=& 2^{k+1}\, \pi\, \ri^k\, \re^{{{\rm i}\pi\over 2} nk^2}\
\prod_{j=1}^{k-1}\sin(\pi n j)\ \times\nonumber  \\
&&
\cos\big(\pi (a Q+
{\textstyle{nk\over 2}})\big)\
\lim_{\varepsilon\to 0} \varepsilon\langle \, I_{2k}(-\ldots+)\,
\rangle|_{\zeta=0}\ .
\eea
Furthermore, it is easy to see that
\bea\label{kajshks}
&&
\lim_{\varepsilon\to 0} \varepsilon\langle \, I_{2k}(-\ldots+)\, \rangle|_{\zeta=0}
=
{\textstyle {1\over k}}\times\\
&& \int_{0}^1\rd z_{2k-1}\ldots\int_{0}^{z_3}{\rd z}_2
\int^{z_{2}}_0{\rd z}_{1}
\ \langle\, V_-(1)  V_+(z_{2k-1})\ldots V_+(z_{1}) V_\lambda(0)\, \rangle
\ ,\nonumber
\eea
and hence
\bea\label{sssksa}
&&G_k={ 2^{k+1}\, \pi\over k}\
\prod_{j=1}^{k-1}\sin(\pi n j)\ \  \cos\big(\pi (a Q+
{\textstyle{nk\over 2}})\big)\times\\
&&\int_{0}^{1}\rd z_k\int_{0}^{z_k}\rd\zeta_{k-1}\ldots
\int_{0}^{\zeta_{1}} \rd z_1\
\prod_{j=1}^{k-1} \Big[\zeta_j^{{kn-1\over 2}-a Q}\, (1-\zeta_j)\Big]\
\prod_{j>i}(\zeta_j-\zeta_i)\times\nonumber \\
&&\prod_{j=1}^{k} \Big[z_j^{{kn-1\over 2}+a Q}\, (1-z_j)^{-n-1}\Big]
\ \prod_{j>i}(z_j-z_i)\
\prod_{j,i}|z_i-\zeta_j|^{-n-1}\ .
\nonumber
\eea

\subsection{Final step of the calculation}

Integrations over the variables $z_1,\,\ldots\, z_k$
in \eqref{sssksa} can be eliminated by using
the following identity\footnote{
This identity can be interpreted  as the Dotsenko-Fateev
representation\ \cite{Dot} of   the simple conformal block
of Virasoro algebra with the central charge $c=-2$.
A similar  identity was
used in Ref.\cite{fata}.}:
\bea\label{aslsas} &&
\int_{\zeta_{k-1}}^{\zeta_k}
\rd z_k \int_{\zeta_{k-1}}^{\zeta_{k-2}}\rd z_{k-1}\ldots
\int_{\zeta_{0}}^{\zeta_{1}}\rd z_{1}\  \prod_{j>i} (z_j-z_i)\times\\
&& \prod_{j=0}^{k} \prod_{i=1}^k |\zeta_j-z_i|^{\alpha_j}=
 {\prod_{j=0}^{k}\Gamma(1+\alpha_j)\over
\Gamma(k+1+\sum_{j=0}^k \alpha_j )}\ \ \prod_{j>i} (\zeta_j-\zeta_i)^{\alpha_i+
\alpha_j+1}
\, ,\nonumber
\eea
where $\zeta_k>\zeta_{k-1}>\ldots>\zeta_0$ is an ordered set of real numbers.
In the case under consideration $\zeta_0=0,\ \zeta_{k}=1$ and
\bea\label{slsakjlk}
\alpha_0={kn-1\over 2}+a Q\, , \ \  \ \alpha_{j}=-n-1\ \ \ (j=1,\,\ldots\, k) \ .
\eea
Thus
\bea\label{asksjsal}
G_k=
{2^{k+1}\, \pi^2\  D_{k-1}\over k\, \Gamma({1-kn\over 2}-a Q)
\Gamma({1-kn\over 2}+a Q)}\
\Gamma^k(-n)\ \prod_{j=1}^{k-1}\sin(\pi nj)
\ ,
\eea
where $D_{k-1}$ is  the Selberg integral\ \cite{Selberg, Dot}:
\bea\label{salkks}
&&D_{k-1}
=\int_{0}^{1}\rd \zeta_{k-1} \int_{0}^{\zeta_{k-1}}\rd \zeta_{k-2}\ldots
\int_{0}^{\zeta_{2}}\rd \zeta_{1} \prod_{j=1}^{k-1}
\zeta_j^{(k-1)n-1}(1-\zeta_j)^{-2n}\times \nonumber\\ &&
\ \ \ \   \prod_{j>i}(\zeta_j-\zeta_i)^{-2n}=
{ \pi^{k-1}\ (-n)^{-k}\, \Gamma(1-kn)\over (k-1)!\ \Gamma^k(-n)}\ {1\over
\prod_{j=1}^{k-1}\sin(\pi n j )}\ \, .
\eea
Combining Eqs.\eqref{asksjsal} and \eqref{salkks}\ one arrives to\
\eqref{gkanswer}.

\end{document}